# AN ENHANCED STATIC DATA COMPRESSION SCHEME OF BENGALI SHORT MESSAGE

Abu Shamim Mohammad Arif

Assistant Professor,
Computer Science & Engineering Discipline,
Khulna University,
Khulna, Bangladesh
E-mail: shamimarif@yahoo.com

Asif Mahamud

Computer Science & Engineering Discipline,
Khulna University,
Khulna, Bangladesh.
E-mail: asif.cse04@gmail.com

Rashedul Islam

Computer Science & Engineering Discipline,
Khulna University,
Khulna, Bangladesh.
E-mail: rashedrst@yahoo.com

*Abstract*—**This paper concerns a modified approach of compressing Short Bengali Text Message for small devices. The prime objective of this research technique is to establish a low-complexity compression scheme suitable for small devices having small memory and relatively lower processing speed. The basic aim is not to compress text of any size up to its maximum level without having any constraint on space and time; rather than the main target is to compress short messages up to an optimal level which needs minimum space, consume less time and the processor requirement is lower. We have implemented Character Masking, Dictionary Matching, Associative rule of data mining and Hyphenation algorithm for syllable based compression in hierarchical steps to achieve low complexity lossless compression of text message for any mobile devices. The scheme to choose the diagrams are performed on the basis of extensive statistical model and the static Huffman coding is done through the same context.**

## I. INTRODUCTION

We are now at the age of science. Now a day, Science brings everything to the door of us. Science makes life easy with its many renowned and unrenowned achievements. Small devices are one of such achievements. In case of our personal computer there is much space to store various types of data. We never worried about how much space the data or messages take into the memory to store that data. But in case of small device we have to consider the memory space required to store the respective data or text messages. Compression of the text message is the number one technique in this case.

Compression is an art of reducing the size of a file by removing redundancy in its structure. Data Compression offers an attractive approach of reducing communication costs by using available bandwidth effectively. Data Compression technique can be divided into two main categories namely for Lossless Data Compression and Lossy Data Compression. If the recovery of data is exact then the compression algorithms are called Lossless. This type of lossless compression algorithms are used for all kinds of text, scientific and statistical databases, medical and biological images and so on. The main usage of Lossy Data Compression is in normal image compression and in multimedia compression. Our aim is to develop a Lossless Compression technique for compressing short message for small devices.

It is necessary to clearly mention here that compression for small devices may not be the ultimate and maximum compression. It is because of the case that in order to ensure compression in the maximum level we definitely need to use and implement algorithms sacrificing space and time. But these two are the basic limitations for any kind of mobile devices especially cellular phones. Thus we are to be concerned on such techniques suitable to compress data in the most smart and efficient way from the point of view of low space and relatively slower performance facility and which is not require higher processor configuration.

The basic objective of the thesis is to implement a compression technique suitable for small devices to facilitate to store text messages by compressing it up to a certain level. More precisely saying- Firstly, to achieve a technique which is simple and better to store data in a small device. Secondly, to keep required compression space minimum in order to cope with memory of small devices. Thirdly, to have the compression time optimal and sustainable.

## II. LITERATURE SURVEY

### A. Definitions

*Data Compression*

In computer science and information theory, data compression often referred to as source coding is the process of encoding information using fewer bits (or other information-bearing units) than an un-encoded representation would use through use of specific encoding schemes. One popular instance of compression that many computer users are familiar with is the ZIP file format, which, as well as providing compression, acts as an achiever, storing many files in a single output file.

As is the case with any form of communication, compressed data communication only works when both the sender and receiver of the information understand the encoding scheme. For example, this text makes sense only if the receiver understands that it is intended to be interpreted as characters representing the English language. Similarly, compressed data can only be understood if the decoding method is known by the receiver. Some compression algorithms exploit this property in order to encrypt data during the compression process so that decompression can only be achieved by an authorized party (eg. through the use of a password). [9]

Compression is useful because it helps reduce the consumption of expensive resources, such as disk space or transmission bandwidth. On the downside, compressed data must be uncompressed to be viewed (or heard), and this extra processing may be detrimental to some applications. For instance, a compression scheme for text requires mechanism for the text to be decompressed fast enough to be viewed as it's being decompressed and may even require extra temporary space to decompress the text. The design of data compression schemes therefore involve trade-offs between various factors, including the degree of compression, the amount of distortion introduced (if using a lossy compression scheme), and the computational resources required to compress and uncompress the data.

*Short Message*

A message in its most general meaning is an object of communication. It is something which provides information; it can also be this information itself [9]. Therefore, its meaning is dependent upon the context in which it is used; the term may apply to both the information and its form. A communiqué is a brief report or statement released by a public agency. [9]

*Short Text Message*

Text Messaging, also called SMS (Short Message Service) allows short text messages to be received and displayed on the phone. 2-Way Text Messaging, also called MO-SMS (Mobile-Originated Short Message Service,) allows messages to be sent from the phone as well.[9] Text messaging implies sending short messages generally no more than a couple of hundred characters in length. The term is usually applied to messaging that takes place between two or more mobile devices

*Existing Methods and Systems for Lossless Data Compression*

Though a number of researches have been performed regarding data compression, in the specific field of SMS Compression the number of available research works is not huge. The remarkable subject is that all the compression technique is for other languages but not for Bengali. The techniques are mainly for English, Chinese, and Arabic etc. Bengali differs from these languages for its distinct symbol and conjunct letters. So, we have to gather knowledge from the other language compression technique and then had to go for our respective compression. The following two sections give a glimpse of the most recent research developments on SMS Compression issue.

*Efficient Data Compression Scheme using Dynamic Huffman Code Applied on Arabic Language [1]*

This method is proposed by Sameh *et al*. In addition to the categorization of data compression schemes with respect to message and codeword lengths, these methods are classified as either static or dynamic. A static method is one in which the mapping from the set of messages to the set of code-words is fixed before transmission begins, so that a given message is represented by the same codeword every time it appears in the message ensemble. The classic static defined-word scheme is Huffman coding. In Huffman coding, the assignment of code-words to source messages is based on the probabilities with which the source messages appear in the message ensemble. Messages which appear more frequently are represented by short code-words; messages with smaller probabilities map to longer code-words. These probabilities are determined before transmission begins. A code is dynamic if the mapping from the set of messages to the set of code-words changes over time. For example, dynamic Huffman coding involves computing an approximation to the probabilities of occurrence "on the fly", as the ensemble is being transmitted. The assignment of code-words to messages is based on the values of the relative frequencies of occurrence at each point in time. A message x may be represented by a short codeword early in the transmission because it occurs frequently at the beginning of the ensemble, even though its probability of occurrence over the total ensemble is low. Later, when the more probable messages begin to occur with higher frequency, the short codeword will be mapped to one of the higher probability messages and x will be mapped to a longer codeword. There are two methods to represent data before transmission: Fixed Length Code and Variable length Code.

The Huffman coding algorithm produces an optimal variable length prefix code for a given alphabet in which frequencies are pre assigned to each letter in the alphabet. Symbols that occur more frequently have shorter Code words than symbols that occur less frequently. The two symbols that occur least frequently will have the same codeword length. Entropy is a measure of the information content of data. The entropy of the data will specify the amount of lossless data compression can be achieved. However, finding the entropy of data sets is non trivial. We have to notice that there is no unique Huffman code because Assigning 0 and 1 to the branches is arbitrary and if there are more nodes with the same probability, it doesn't matter how they are connected

The average message length as a measure of efficiency of the code has been adopted in this work.

The average search length of the massage is the summation of the multiplication of the length of code-word and its probability of occurrence.

Also the compression ratio as a measure of efficiency has been used.

Comp. Ratio = Compressed file size / source file size * 100 %

The task of compression consists of two components, an encoding algorithm that takes a message and generates a "compressed" representation (hopefully with fewer bits) and a decoding algorithm that reconstructs the original message or some approximation of it from the compressed representation.

*Genetic Algorithms in Syllable-Based Text Compression [2]*

This method is proposed by Tomas Kuthan and Jan Lansky. To perform syllable-based compression, a procedure is needed for decomposition into syllables. They call an algorithm hyphenation algorithm if, whenever given a word of a language, it returns it's decomposition into syllables. According to the definition of syllable every two different hyphenation of the same word always contain the same number of syllables. There can be an algorithm that works as a hyphenation algorithm for every language. Then it is called universal hyphenation algorithm. Otherwise it is called specific hyphenation algorithm. They describe four universal hyphenation algorithms: universal left $P_{UL}$, universal right $P_{UR}$, universal middle-left $P_{UML}$ and universal middle-right $P_{UMR}$.

The first phase of all these algorithms is the same. Firstly, they decompose the given text into words and for each word mark its consonants and vowels. Then all the maximal subsequences of vowel are determined. These blocks form the ground of the syllables. All the consonants before the first block belong to the first syllable and those behind the last block will belong to the last syllable.

This algorithm differs in the way they redistribute the inner groups of consonants between the two adjusting vowel blocks. $P_{UL}$ puts all the consonants to the preceding block and $P_{UR}$ puts them all to the subsequent block. $P_{UML}$ and $P_{UMR}$ try to redistribute the consonant block equally. If their number is odd $P_{UML}$ pushes the bigger parity to the left, while $P_{UMR}$ to the right. The only exception is, when $P_{UML}$ deals with a one-element group of consonants. It puts the only consonant to the right to avoid creation of not so common syllables beginning with a vowel.

**Hyphenating priesthood**

| correct hyphenation | priest-hood |
| universal left $P_{UL}$ | priesth-ood |
| universal right $P_{UR}$ | prie-sthood |
| universal middle-left $P_{UML}$ | priest-hood |
| universal middle-right $P_{UMR}$ | pries-thood |

Effectiveness of these algorithms is then measured. In general, $P_{UL}$ was the worst one; it had lowest number of correct hyphenations and produced largest sets of unique syllables. The main reason for this is that it generates a lot of vowel-started syllables, which are not very common. $P_{UR}$ was better but the most successful were both 'middle' versions. English documents were best hyphenated by $P_{UMR}$, while with Czech texts $P_{UML}$ was slightly better.

*Lossless Compression of Short English Text Message for JAVA Enable Mobile Devices [3]*

This method is proposed by Rafiqul *et al.* published in preceedings of 11th International Conference on Computer and Information Technology (ICCIT 2008), in December, 2008, Dhaka, Bangladesh. The Total compression process is divided into three steps namely for Character Masking, Substring Substitution or Dictionary Matching or Partial Coding, Bit Mapping or Encoding (Using Fixed Modified Huffman Coding).

The very first step of planned SMS Compression is Character Masking. Character Masking is a process in which the character(s) code(s) are changed or re-defined on the basis of any specific criteria. In concerned task it is planned to use character masking for reducing the storage overhead for blank spaces. Firstly the spaces are expected to be searched out and then encoded by a predefined code-word. This codeword should be unique for the overall compression. For multiple consecutive blank spaces the same technique may be employed. The modified message is then passed towards the next compression step for dictionary Matching or Partial Coding.

In the second step they employ Dictionary Matching or Partial Coding. In this phase the string referred from the first step is passed through a partial coding which encodes the masked characters (performed previously in the first step) on the basis of the following character. The character following masked character mergers the masked space by encoding it. Thus all the spaces are merged and as a result it may be certainly reduce a mentionable amount of characters. After this task we pass the modified string of message through a dictionary matching scheme where the message is searched for matching some pre-defined most-commonly used words or substrings or punctuations to reduce the total number of characters. The message is then forwarded to Step 3 where the actual coding is performed.

In the final step of coding they have used static Huffman Coding Style. But here the modification is made that in spite of calculating on-stage codes they use predefined codes in order to reduce the space and time complexity. The codes for dictionary entries are also predefined. The total message is thus encoded by a comparatively small number of bits and hence they get a compressed outcome.

*Compression of a Chinese Text [4]*

This method is proposed by Phil Vines, Justin Zobel. In this method, the byte oriented version of PPM (Partial Predictive Matching) is not predicting characters, but rather halves of character. It is reasonable to suppose that modifying PPM to deal with 16 bit characters should enable the model to more accurately capture the structure of the language and hence provide better compression .They have identified several changes that need to be made to the PPM implementation described above to allow effective 16-bit coding of Chinese. First, the halving limit needs to be modified the number of 16-bit characters that can be occur in a context in much greater than of 4-bit characters. So a large probability space is required. Second, in conjunction with this change the increment should also be increased to force more frequent halving and prevent the model from stagnating. Their experiments suggest that a halving limit of 1024 and an increment of 16 are appropriate. Third, the method described above for estimating escape probabilities may not be appropriate since so many characters are novel. Fourth, model order must be chosen.

Most implements encode bytes, but this is an arbitrary choice and any unit can be used within the constraints of memory size and model order. For English contexts and symbols are quickly repeated, so that, after only a few kilobytes of text, good compression is achieved and contexts of as little as three characters can give excellent compression.

As byte-oriented PPM is a general method that gives good results not only for English text but for a wide variety of data types, an obvious option is to apply it directly to Chinese text. Higher order models take a loner time to accumulate contexts with probabilities that accurately reflect the distribution. So that, when memory is limited, the models spends most of its time in the learning phase, where it emits large number of escape codes and is unable to make . Thus they observe poorer compression because such contexts do not reappear sufficiently often before the model needs to be flushed and rebuild over 800. Reloading the model with immediate prior text after each flush is unlikely to be helpful; since the problem is that there is not sufficient memory to hold the model that make accurate prediction. It follows that increasing the amount of memory available for storing contexts could be expected to improve compression performance. However, assuming only moderate volumes of memory are available, managing even a character-based model can be problematic; they believe that because of the number of distinct symbols, use of a word-based model is unlikely to be valuable. The implementation of PPM described above uses a simple memory management strategy; all information is discarded when the available space is consumed.

### III. PROPOSED SYSTEM

Our prime concern of thesis is to implement a lossless compression of short Bengali text for low-powered devices in a low complexity scheme. The idea behind this is there are still many compression techniques for languages like English, Arabic and other language and many people are still involving to improve the compression ratio of messages of the respective language. Some of them are also published in various conferences and journals. Although Bengali short message technique is achieved couple of years ago but there is not still any compression technique suitable for Bengali languages.

Bengali text compression differs from English text compression from mainly two points of views. Firstly, the compression techniques involving pseudo-coding of uppercase (or lowercase) letters are not applicable for Bengali text. Secondly, in case of Bengali, we may employ specific mechanism of coding dependent vowel signs to remove redundancy, which is absent for the case of English. In Bengali, we have 91 distinct symbol units including independent vowels, constants, dependent vowel signs, two part independent vowel signs, additional constants, various signs, additional signs and Bengali numerals etc. A detail of Bengali symbols available in. Moreover, in Bengali we have a large involvement of conjuncts which also focuses a scope of redundancy removal.

Though English has got a fixed encoding base long ago, still now in practical application, Bengali has adapted unique encoding scheme. The use of Bengali Unicode has not yet got a massive use. This is really a great limitation for research in Bengali. Bengali text compression also suffers from the same problem.

#### A. *Compression Process*

In this paper, we propose a new dictionary based compression technique for Bengali text compression. To facilitate efficient searching and low complex coding of the source text, we employ term probability of occurring characters and group of characters in a message with indexing the dictionary entries. The total compression scheme is divided into two stages:

Stage 1: Building the knowledge-base.

Stage 2: Apply proposed text ranking approach for compression the source text.

### Stage 1: Building the knowledge-base

The test-bed is formed from the standard Bengali text collections from various sources. We consider a collection of texts of various categories and themes (like news, documents, papers, essays, poems and advertising documents) as the test bed. By reading the respective frequency statistics we select our respective knowledgebase entries and divide the frequency into a four level architecture. Assigning minimum length code-words to the selected components is the main objective of the statistics gathering phase. It is remarkable that, though a few collections of domain specific text collection are available, still now no sophisticated Bengali text compression evaluation test-bed is available. As data compression and especially dictionary based text compression greatly involves the structure, wording and context of texts, a collection involving different types of text is a must for evaluating the compression. In constructing the dictionary, we use the test-text-bed of 109 files varying from 4kb to 1800kb.

The Variable Length Coding (VLC) algorithm [1] is used to produce an optimal variable length prefix code for a given alphabet. Noteworthy that, in the previous step of knowledgebase formation, frequencies is already pre-assigned to each letter in the alphabet. Symbols that occur more frequently have shorter Code-words than symbols that occur less frequently. The two symbols that occur least frequently will have the same codeword length. Entropy is a measure of the information content of data. The entropy of the data will specify the amount of lossless data compression can be achieved. However, finding the entropy of data sets is non trivial. We have to notice that there is no unique Huffman code because Assigning 0 and 1 to the branches is arbitrary and if there are more nodes with the same probability, it doesn't matter how they are connected.

The average message length as a measure of efficiency of the code has been adopted in this work.

Avg L = L1 * P (1) + L2 * P (2) + ….. + Li * P (i)

Avg L = $\sum$ Li * P (i)

Also the compression ratio as a measure of efficiency has been used.

Compression Ratio = Compressed file size / source file size * 100 %

The task of compression consists of two components, an encoding algorithm that takes a message and generates a "compressed" representation (hopefully with fewer bits) and a decoding algorithm that reconstructs the original message or some approximation of it from the compressed representation.

**Stage 2: Apply proposed text ranking approach for compression the source text**

Text ranking is an elementary scheme which is used to assign weights or index of texts or terms (especially word tokens) on the basis of any suitable scheme or criteria. This scheme of indexing or ranking is ideally frequency of occurrence of the texts or even probability of occurrence of the texts or components. In our method, we grab the source text and take the Unicode value of the corresponding data. Our method of compression process differs mainly from others in this point. Still now no one has the method of taking the most successive match. But we have the way to take the most successive match. We will start with maximum level and proceed through the hierarchical levels to find successful match. It is to remark that in the last level there is only letters and their Unicode value. So, if a word does not match in any level it has to match in this level. To perform compression, we need a procedure for decomposition into syllables. We will call an algorithm hyphenation algorithm [2] if, whenever given a word of a language, it returns it's decomposition into syllables. It is called universal hyphenation algorithm. By this algorithm we can generate the successful match for a string or sentence what other method haven't done. We call this algorithm as specific hyphenation algorithm. We will use four universal hyphenation algorithms:

universal left $P_{UL}$,

universal right $P_{UR}$,

universal middle-left $P_{UML}$ and

universal middle-right $P_{UMR}$.

The first phase of all these algorithms is the same. Firstly, we decompose the given text into words and for each word mark its letter and symbol. Then we determine all the maximal subsequences of vowel. These blocks form the ground of the syllables. All the consonants before the first block belong to the first syllable and those behind the last block will belong to the last syllable.

After taking the most successive matching then encode with the code-words obtained in the Step 1 for each matching elements. And lastly the resultant data will be transmitted.

*B. Decompression Process*

The total decompression process can be divided into the following three steps:

**Step 1:** Grab the bit representation of the message

**Step 2:** Identify the character representation

**Step 3:** Display the decoded message.

As all the letters and symbols are to be coded in such a fashion that by looking ahead several symbols (Typically the maximum length of the code) we can distinguish each character (with attribute of Human Coding).

In step 1 the bit representation of the modified message is performed. It is simply analyzing the bitmaps.

The second step involves recognition of each separate bit-patterns and indication of the characters or symbols indicated by each bit pattern. This recognition is performed on the basis of the information from fixed encoding table used at the time of encoding.

The final step involves simply representing i.e. display of the characters recognized through decoding the received encoded message.

### IV. EXPERIMENT RESULT

The proposed model of short text Compression for Bengali language provides much more efficiency than other SMS compression models. This proposed model is also expected to have lower complexity than that of the remaining models. The steps provided are not previously implemented in same model. The basic aspect of the model is, in this model we plan to use less than eight bit codeword for each character in average using static coding in place of eight bits and hence we may easily reduce total number of bits required in general to represent or transmit the message. The modification is required and to some extent essential because for low complexity devices is not any intelligent approach to calculate the codes for each character sacrificing time and space requirements. That is why it may be a good approach to predefine the codes for each character having less bit length in total to compress the message. The fixed codes will be determined from the heuristic values based on the dictionary we normally use. The ultimate aim is to use less number of bits to reduce the load of characters.

We intend to apply the dictionary matching or multi-grams method to enhance the optimality of compression. Multi-grams method is used in order to replace a number of used sub-string or even strings from the input message. Specific code words are defined for those words or substrings or strings. It is because the case that if we can replace any part of the message by a single characters then we can definitely reduce the total number of character gradually. It is necessary to mention here that the co-ordination of multi-grams or dictionary method with modified Huffman coding may ensure the maximum 3 to 5 possible compression. In order to enhance the performance of compression the dictionary matching or multi-grams will play a vital role in compression ratio because the propose thesis is based on successful and ultimate optimal compression of Bengali text at the level best for wireless mobile devices with small memory and lower performance speed. As we are using Huffman coding for length seven whereas each character requires eight bits to be represented. Thus for n characters we will be able to compress n bits using fixed Huffman coding. In the next step we will be able to save the memory requirements for blank spaces using character masking. For example, for any Bengali short message of length 200 characters it is usual to predict that we may have at least 25 spaces. If we can eliminate those blank spaces by masking with its consecutive character through character masking, then we may reduce those 25 characters from the original message. It is necessary to mention here that the dictionary matching or multi-grams method is completely dependent on the probability distribution of input message and we are to be much more careful on choosing the dictionary entries for Bengali text.

The implementation of the compression scheme is performed using JAVA 2 Micro Edition with the simulation using Java-2 SE 1.5. The cellular phones adapting the proposed Compression tool must be JAVA powered. The implementation will include both encoding and decoding mechanism.

A. *Discussions on Results*

The performance evaluation is performed on the basis of the various corpuses. As the prime aspect of our proposed Compression Scheme is not to compress huge amount of text rather to compress texts with limited size affordable by 36 the mobile devices i.e. embedded systems, we took blocks of texts less than one thousand characters chosen randomly from those files ignoring binary files and other Corpus files and performed the efficiency evaluation.

The most recent study involving compression of text data are-

1. "Arabic Text Steganography using multiple diacritics" by Adnan Abdul-Aziz Gutub, Yousef Salem Elarian, Sameh Mohammad Awaideh, Aleem Khalid Alvi. [1]

2. "Lossless Compression of Short English Text Message for JAVA enables mobile devices" by "Md. Rafiqul Islam, S. A. Ahsan Rajon, Anondo Poddar. [2]

We denote the above two methods as DCM-1 and DCM-2 respectively.

The simulation was performed in a 2.0 GHz Personal Computer with 128 MB of RAM in threading enable platform. The result for different size of blocks of text is as follows-

| Source | *DCM-1* | *DCM-2* | *Proposed Technique* |
|---|---|---|---|
| Prothom Alo | 4.24 | 4.01 | 3.98 |
| Vorer Kagoj | 3.78 | 4.19 | 3.98 |
| Amader somoy | 4.02 | 4.08 | 3.93 |
| Ekushe-khul.poem | 4.98 | 3.98 | 3.65 |
| Ekushe-khul.Article | 4.48 | 3.79 | 3.44 |

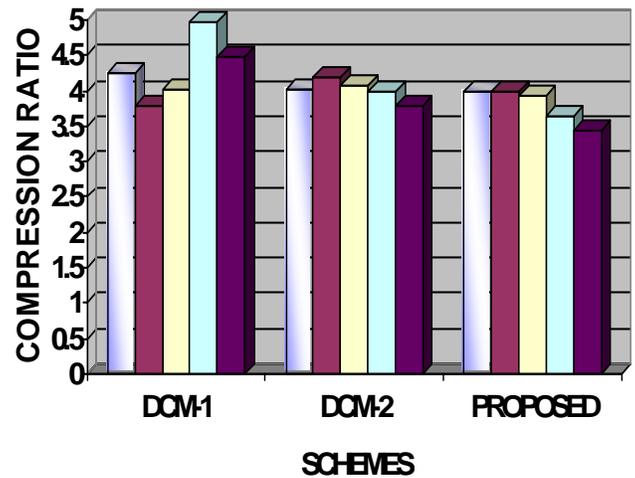

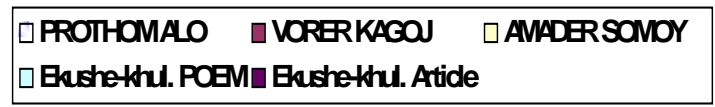

V. CONCLUSION

The prime objective of this undergraduate research is to develop a more convenient low complexity compression technique for small devices. As the environment is completely different from the usual one (PCs with huge memory and amazingly greater performance speed) and the challenge is to cope with the low memory and relatively less processing speed of the cellular phones, the ultimate objective is to devise a way to compress text messages in a smart fashion to ensure optimal rate and efficiency for the mobile phones which may not be the best approach for other large-scale computing devices. That is why, in comparison to other ordinary data compression schemes the proposed is of low complexity and less time consuming.